\documentclass[12pt,a4paper]{article}

\usepackage{amsmath}
\usepackage{amssymb}
\usepackage{graphicx}
\usepackage{bm}

\newcommand\be{\begin{equation}}
\newcommand\ee{\end{equation}}
\newcommand\non{\nonumber}
\newcommand\eps{\epsilon}
\newcommand\half{\frac{1}{2}}
\newcommand\supp{\mathop{\rm supp}\nolimits}
\def\RR{\mathbb{R}}
\def\ZZ{\mathbb{Z}}

\def \outlineby #1#2#3{\vbox{\hrule\hbox{\vrule\kern #1%
\vbox{\kern #2 #3\kern #2}\kern #1\vrule}\hrule}}%
\def \endbox {\outlineby{4pt}{4pt}{}}%
\newcommand{\qed}{\hfill \endbox}

\def\tcdms{T. C. Dorlas and M. Samsonov\\
{\it Dublin Institute for Advanced Studies}\\
{\it School of Theoretical Physics}\\
{\it 10 Burlington Road, Dublin 4, Ireland.}}

\newtheorem{theorem}{Theorem}[section]
\newtheorem{lemma}{Lemma}[section]
\newtheorem{prop}{Proposition}[section]
\newtheorem{corollary}{Corollary}[section]

\def\sgn{{\rm \sgn}}
\def\shalf{{\scriptstyle{\frac{1}{2}}}}

\begin{document}

\title{On the thermodynamic limit of the 6-vertex model}
\author{\tcdms}

\maketitle

\begin{abstract} 
We give a rigorous treatment to the thermodynamic limit 
of the 6-vertex model. We prove that the unique solution of the Bethe-Ansatz equation exists and the distribution of the roots converges to a continuum measure. We solve this problem for $0<\Delta<1$ using convexity arguments and for large negative $\Delta$ using the Fixed Point Theory of appropriately defined contracting operator.
 \end{abstract}


\section{The 6-vertex model and formulation of the problem}

The 6-vertex model is an exactly soluble model of classical
statistical mechanics introduced and solved in various special
cases by Lieb \cite{Lieb1, Lieb2, Lieb3}. A solution of the most
general case was obtained by Sutherland \cite{Su}. A clear
description of this model and various other soluble models can be
found in Baxter's book \cite{Baxter}. However, as Baxter remarks,
an exact solution is not the same as a rigorous solution. In fact,
already in his first article on the ice model \cite{Lieb1}, Lieb
initiated the rigorous analysis of the model. A more extensive
analysis was made by Lieb and Wu \cite{LiebWu}. An important
technical question was left unresolved, however. This concerns the
convergence of the distribution of (quasi-) wavenumbers to a
continuum measure in the thermodynamic limit. (Another technical
problem, i.e. the independence of the free energy on the boundary
conditions, was resolved by Brascamp et al. \cite{Brascamp}.) A
similar problem was solved in the case of the nonlinear
Schroedinger model in \cite{DLP}. The 6-vertex model is more
complicated because we cannot in all cases use the convexity
argument of Yang and Yang \cite{YY1} used there. However, their
argument does extend to a certain domain of parameter space. Here
we show how it can be used to prove the convergence of the Bethe
Ansatz solutions in the thermodynamic limit in that case. In
addition, we use another technique for proving the existence of a
unique solution to the Bethe Ansatz equations in the thermodynamic
limit in a different domain of parameter space. Uniqueness in
other parts of parameter space is still an open problem, though
numerical iteration does seem to converge to a unique solution.

\subsection{Definition of the model and the free energy}

We first recall the definition of the 6-vertex model and some
general results concerning  the existence of the thermodynamic
limit. We then review the transfer matrix formulation of the model
and the diagonalisation of the transfer matrix by means of the
Bethe Ansatz.

The 6-vertex model is a model of classical statistical mechanics
where the configurations are given by arrows on the bonds of a
2-dimensional square lattice. At each vertex only six different
configurations of arrows are allowed (the so-called ice condition): 

\includegraphics[angle=0,width=12cm]{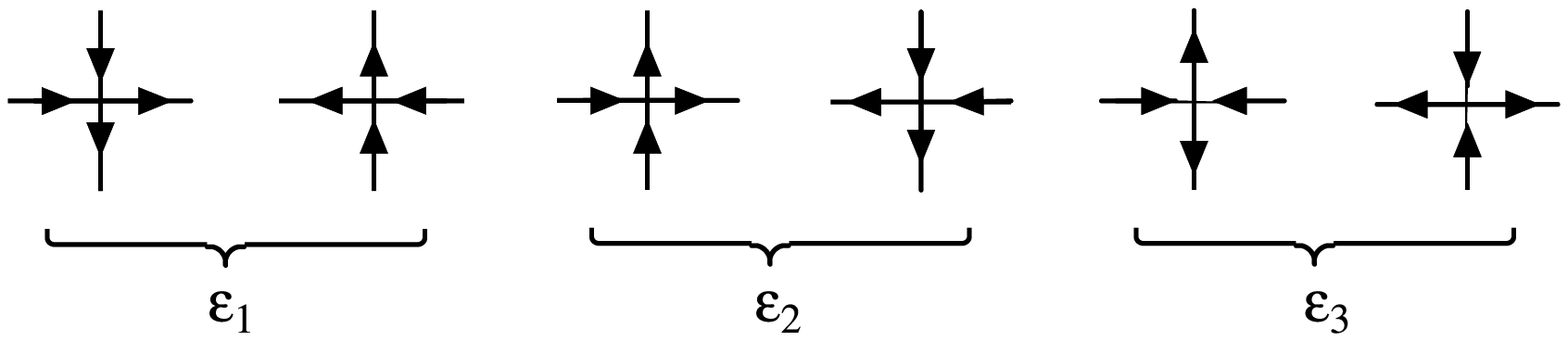}

Each of these vertex configurations is assigned an energy and we
assume spin-flip invariance, so that the first and the second, the
third and the fourth and the fifth and the sixth configuration
have the same energy. We denote these energies by  $\eps_1,\
\eps_2$ and $\eps_3.$ If $\beta$ is the inverse temperature, the
corresponding Boltzmann weights are: $a=\exp[-\beta\eps_1],\quad
b=\exp[-\beta\eps_2]$ and $c=\exp[-\beta\eps_3].$ The partition
function is therefore  \be \label{partfn} Z_{M,N} (a,b,c) =
\sum_{\Gamma \in {\cal C}_{M,N}} e^{-\beta E(\Gamma)}, \ee where
$M$ is the number of rows and $N$ is the number of columns in the
lattice, ${\cal C}_{M,N}$ denotes the set of allowed
configurations, and the total energy of a configuration $\Gamma
\in {\cal C}_{M,N}$ is \be E(\Gamma) = n_1(\Gamma)\, \eps_1 +
n_2(\Gamma)\, \eps_2 + n_3(\Gamma)\, \eps_3, \ee if $n_i(\Gamma)$
is the number of vertices of type $i$ in the configuration
$\Gamma$. \lq\lq Solving this model'' now means: finding an
explicit expression for the thermodynamic limit of the free energy
density, i.e. \be \label{free-energy}
f(\eps_1,\eps_2,\eps_3;\beta) = -\frac{1}{\beta} \lim_{N,M \to
\infty} \frac{1}{NM} \ln Z_{N,M} (a,b,c) \ee We shall assume
periodic boundary conditions.

The first question that arises is whether the limit
(\ref{free-energy}) exists. This was solved by Lieb and Wu
\cite{LiebWu}. In fact, we need the existence of the free energy
at constant density $\rho$. In \cite{LiebWu} this is highlighted
as an open problem, but in fact, in the case of periodic boundary
conditions, their method extends to this case. For convenience we
repeat their argument here.  (It was proved by Brascamp et.al.
\cite{Brascamp} that periodic boundary conditions are equivalent
to free boundary conditions in the thermodynamic limit. It should
be noted that not all boundary conditions are equivalent: see the
recent solution of the model with domain-wall boundary conditions
by Bleher et al. \cite{Bleher}.)

The periodic boundary conditions imply that, in a given
configuration, the number of up arrows in every row of vertical
arrows is the same. We shall call this number divided by the
maximum number $N$, the \textit{density} $\rho$. The partition
function with fixed density $\rho$ is given by \be \label{pfd}
Z_{M,N} (w_1,w_2,w_3;\rho) =
\sum_{\alpha:\,\#\{\alpha_i=+1\}=N\rho} \sum_\gamma \sum_{\Gamma
\in {\cal C}^p_{M,N}(\alpha,\gamma)} e^{-\beta E(\Gamma)}, \ee
where ${\cal C}_{M,N}^p (\alpha,\gamma) = {\cal C}_{M,N}
(\alpha,\alpha,\gamma,\gamma)$ and ${\cal C}_{M,N}
(\alpha,\alpha^\prime,\gamma,\gamma^\prime)$ denotes the set of
configurations with given boundary arrows: $\alpha$ and
$\alpha^\prime$ for the bottom and top rows of vertical arrows,
and $\gamma$ and $\gamma^\prime$ for the left- and right-hand
columns of horizontal arrows.

\begin{prop} Let $Z^p_{M,N}$ denote the partition function of the
six-vertex model with periodic boundary conditions and let
$(M_l,N_l)$ be a sequence tending to infinity in the sense of
Van Hove, and suppose that $(\rho_l)_{l=1}^\infty$ is a sequence
of numbers $\rho_l \in [0,1]$ tending to $\rho$ such that
$\rho_l N_l \in \mathbb{N}$. Then the corresponding free energy
density $f^p(\eps_1,\eps_2,\eps_3;\beta,\rho)$ defined by \be
\label{fed} f^p (\eps_1,\eps_2,\eps_3;\beta,\rho) =
-\frac{1}{\beta} \lim_{l \to \infty} \frac{1}{M_lN_l}
\ln\,Z_{M_l,N_l}^p (a,b,c;\rho_l) \ee exists and is independent
of
the sequences $(M_l,N_l)$ and $(\rho_l)$. Moreover, 
$f^p(\eps_1,\eps_2,\eps_3;\beta,\rho)$ is convex as a function of
$\rho$ and concave as a function of the variables
$\eps_1,\eps_2,\eps_3$ and $\beta$. \end{prop}

{\it Proof.} We start by considering special sequences. Assume
first that $\rho \in [0,1] \cap \mathbb{Q}.$ Take $N_0 \in
\mathbb{N}$ so large that $\rho N_0 \in \mathbb{N}$, and choose
$M_0 \in \mathbb{N}$ arbitrary. Consider a {\it standard sequence}
of rectangular boxes of height $M_l = 2^l M_0$ and width $N_l =
2^l N_0$.  One then proves as in Lieb and Wu \cite{LiebWu} that
the limit (\ref{fed}) exists, using the inequalities
\begin{eqnarray} \label{zbound}  Z_{M_l,N_l} (\beta,\rho)
\geq \Theta_{M_l,N_l} (\beta,\rho) & \geq & \left(
\Theta_{M_{l-1},N_{l-1}} (\beta,\rho) \right)^4 \non \\ & \geq &
\left( Z_{M_{l-1},N_{l-1}} (\beta,\rho) \right) 2^{-4(M_0+N_0)}.
\end{eqnarray} (We suppress the dependence on $\eps_i$ and on the periodic
boundary conditions.) Here we define \be \Theta_{M,N}(\beta,\rho)
= \max_{\alpha:\,\#\{\alpha_i=+1\}=N\rho} \max_{\gamma}
\sum_{\Gamma \in {\cal C}^p_{M,N}(\alpha,\gamma)} e^{-\beta
E(\Gamma)}. \ee The inequalities (\ref{zbound}) imply that the
sequence \be f_{M_l,N_l} (\beta,\rho) = -\frac{1}{\beta M_l N_l}
\ln Z_{M_l,N_l} (\beta,\rho) \ee  is essentially decreasing. As it
is also bounded below, it converges. The concavity as a function
of $\eps_i$ is standard. To prove the convexity as a function of
the density, suppose that $\rho_1 < \rho_2$. Then if $\alpha_1$
and $\alpha_2$ are given sets of $N$ vertical arrows with
densities $\rho_1$ and $\rho_2$ respectively, and we write $\alpha
= \alpha_1 \cup \alpha_2$ for the union,
\begin{equation} \sum_\gamma \sum_{\Gamma
\in {\cal C}^p_{M,2N}(\alpha,\gamma)} e^{-\beta E(\Gamma)} \geq
\sum_\gamma \sum_{\Gamma_1 \in {\cal C}^p_{M,N}(\alpha_1,\gamma)}
\sum_{\Gamma_2 \in {\cal C}^p_{M,N}(\alpha_2,\gamma)} e^{-\beta
(E(\Gamma_1)+E(\Gamma_2)}.
\end{equation} Summing over $\alpha_1$ and $\alpha_2$ we have
\begin{equation} Z^p_{M,2N} (\shalf(\rho_1+\rho_2)) \geq Z^p_{M,N}
(\rho_1) Z^p_{M,N} (\rho_2) \end{equation} and hence
\begin{equation} f(\beta,\shalf(\rho_1+\rho_2)) \leq \half \left(
f(\beta,\rho_1) + f(\beta,\rho_2) \right). \end{equation}

Convexity implies continuity and we can thus extend the definition
to all $\rho \in (0,1)$.

To show that the definition of $f(\beta,\rho)$ is independent of
$M_0$ and $N_0$ we fill a general domain $\Lambda$ with rectangles
and use the condition $\eps_1 > 0$ and $\eps_2 > 0$ to decorate
the remainder by vertices 1,2,3, and 4, as in \cite{LiebWu}.  \qed

In the exact solution of the six-vertex model one actually takes
the limits $M \to \infty$ and $N \to \infty$ consecutively, but it
was also shown by Lieb and Wu \cite{LiebWu} that, for periodic
boundary conditions, this yields the same limit as (\ref{fed}):

\begin{prop} The double limit
\be \tilde f (\eps_1,\eps_2,\eps_3;\beta,\rho) =
-\frac{1}{\beta} \lim_{N \to \infty} \lim_{M \to \infty}
\frac{1}{NM} \ln Z_{M,N}^p (a,b,c) \ee exists and equals
$f(\eps_1,\eps_2,\eps_3,\beta,\rho)$.
\end{prop}

\subsection{The transfer matrix and its diagonalisation}

The transfer matrix method for solving models of classical
statistical mechanics is common knowledge. Using periodic boundary
conditions one writes the partition function as a trace
\begin{equation} \label{zastrace}
Z_{M,N} = \hbox{ Trace }(V_N^M) \end{equation} where
$V_N$ is the transfer matrix with entries between two rows of
vertical arrows $\alpha$ and $\alpha^\prime$ given by
\begin{equation} \label{transfer}
\left( V_N \right)_{\alpha,\alpha^\prime} = \sum_\gamma
\prod_{n=1}^N \exp\,[-\beta \eps_{\alpha_n}^{\alpha^\prime_n}
(\gamma_n,\gamma_{n+1}) ].
\end{equation}  The sum runs over a row of horizontal arrows
$\gamma=(\gamma_1,\dots,\gamma_N)$, where $\gamma_n$ is the
horizontal arrow between the $(n-1)$-th and the $n$-th vertex. It
follows that  we can take the limit $M \to \infty$ to obtain
\begin{equation} \label{felamax}
f(\beta, \rho) = -\frac{1}{\beta} \lim_{N \to \infty}
\frac{1}{N} \ln \Lambda_{\max} (N),  \end{equation}  where
$\Lambda_{\max} (N) $ is the maximum eigenvalue of the transfer
matrix $V_N$ which exists because $V_N$ satisfies the conditions
of the Perron-Frobenius Theorem. The transfer matrix can be
diagonalised by means of the Bethe Ansatz. If we write
$|x_1,\dots,x_n\rangle$ for the row configuration with $n$
up-arrows then a general wave function in the subspace with $n$
up-arrows can be expressed as
\begin{equation} \label{efn}
\psi = \sum_{1 \leq x_1 < \dots < x_n \leq N} \psi (x_1, \dots,
x_n)\, |x_1, \dots, x_n\rangle. \end{equation}  The Bethe Ansatz
for eigenfunctions of $V_N$ then reads: \begin{equation}
\label{baefn} \psi (x_1, \dots, x_n) = \sum_{\sigma \in {\cal
S}_n} A_\sigma \exp\,[i \sum_{j=1}^n k_{\sigma(j)} x_j].
\end{equation} Here, the sum runs over the set ${\cal S}_n$ of all
permutations of$\{1,\dots,n\}$ and  the coefficients $A_\sigma$
and the wave numbers $k_1,\dots,k_n$ are to be determined by
inserting into the eigenvalue equation. This yields the following
conditions:
\begin{enumerate}
\item The wave numbers must satisfy the simultaneous nonlinear
equations: \begin{equation} \label{bae} e^{iNk_j} = (-1)^{n-1}
\prod_{l=1; l\ne j}^n e^{-i \theta (k_j,k_l)},
\end{equation} where the function $\theta$ is defined by
\begin{equation} \label{theta} \exp \,[-i\theta (k,k^\prime)] = \frac{1 - 2\Delta
e^{ik} + e^{i(k+k^\prime)}}{1 - 2\Delta e^{ik^\prime} +
e^{i(k+k^\prime)} } \end{equation} with \begin{equation}
\label{deltavsw} \Delta = \frac{a^2 + b^2 - c^2}{2ab}.
\end{equation}
\item The corresponding eigenvalue is given by
\begin{equation} \label{evs} \Lambda (k_1, \dots, k_n) = a^N \prod_{j=1}^n
L(e^{ik_j}) + b^N \prod_{j=1}^n M(e^{ik_j}), \end{equation} where
$L(z)$ and $M(z)$ are given by \begin{equation} L(z) = \frac{ab +
(c^2 - b^2)z}{a^2 - ab z}, \end{equation}
\begin{equation} M(z) = \frac{a^2 - c^2 - ab z}{ab - b^2 z}.
\end{equation}
\end{enumerate}

Of course, (\ref{theta}) only defines the function $\theta$ up to
a multiple of $2\pi$. In taking the logarithm of (\ref{bae}), we
shall assume that $-\pi < \theta (k,k^\prime) \leq \pi$. We obtain
\begin{equation} \label{keqn} Nk_j = 2\pi I_j - \sum_{l=1}^n
\theta (k_j,k_l), \end{equation} where $I_j \in \mathbb{Z}$ if
$n$ is odd, and $I_j \in \mathbb{Z} + \half$ if $n$ is even.
These equations are identical to the BA equations found by Bethe
\cite{Bethe} in his solution of the Heisenberg chain. They were
analysed in detail by Yang and Yang \cite{YY1}, who showed that
the ground state of the Heisenberg chain is obtained by choosing
\begin{equation} \label{gs} I_j = j - \half (n+1).
\end{equation} They also showed that, for this choice, the
equations (\ref{keqn}) have a real solution for $k_1, \dots, k_n$.
Lieb \cite{Lieb1} then argued that, as the Heisenberg Hamiltonian
also satisfies the conditions for the Perron-Frobenius Theorem,
the corresponding eigenfunction must be positive, and hence it
must also be the eigenfunction of the transfer matrix with maximum
eigenvalue. We therefore have \begin{equation} \label{lamax}
\Lambda_{\hbox{max}} = \Lambda (k_1, \dots, k_n) \end{equation}
where $k_1, \dots, k_n$ are the solutions of (\ref{keqn}) in case
the $I_j$ are given by (\ref{gs}).

By formula (\ref{felamax}), the free energy is now given by
\begin{equation} \label{BAfe} f(\beta, \rho) = \lim_{N \to \infty}
\min \left\{ \eps_1 - \frac{1}{\beta N} \sum_{j=1}^n \ln L(e^{i
k_j}), \eps_2 - \frac{1}{\beta N} \sum_{j=1}^n \ln M(e^{i k_j})
\right\}.
\end{equation} By the fact that the free energy is convex and
symmetric in the density, the minimum is attained at $\rho=\half$
and the solution for $k_1,\dots,k_n$ corresponding to the integers
(\ref{gs}).

In this paper we address the question of how to compute the
thermodynamic limit (\ref{BAfe}). We want to take the limit $N \to
\infty$, keeping $\rho = n/N$ fixed. One usually makes the
reasonable assumption that, in this limit, the distribution of the
wavenumbers $k_1, \dots, k_n$ tends to a continuous distribution
with density $\rho (k)$. In the following we shall investigate the
validity of this assumption. Following Yang and Yang \cite{YY1},
we consider separately the cases $\Delta \in [0,1)$ and $\Delta <
0$. (The case $\Delta \geq 1$ is trivial.) In the attractive case,
$\Delta \in [0,1)$, we can apply the same reasoning as in the case
of the nonlinear Schroedinger model (see \cite{DLP}) and use the
convexity of a certain functional to prove the existence of a
unique solution to (\ref{keqn}). In the repulsive case, we can
only treat the case $\Delta \ll -1$. The case of smaller negative
$\Delta$ is more delicate. The difficulty is proving the
uniqueness of the solution. We can show, however, that if one
assumes that the solution is monotone, the limiting solution is
unique. Numerical solution of the BA equations seems to suggest
that it is unique even without this assumption, but we have so far
been unable to prove that.

\section{Thermodynamic limit in the case $\Delta \in [0,1)$.}

In taking the thermodynamic limit we distinguish the cases $\Delta
> 1$, $\Delta \in [0,1)$, $\Delta \in (-1,0)$ and $\Delta < -1$.
The case $\Delta > 1$ is trivial (Cf. Baxter \cite{Baxter}) so we
start with the case $\Delta \in [0,1)$. We first prove an analogue
of the existence and uniqueness of a solution to the Bethe Ansatz
equations in the thermodynamic limit. In the present case this is
analogous to the nonlinear Schr\"odinger problem treated in
\cite{DLP}.

\begin{theorem} \label{thm1}
Let $m \in {\cal M}_+^b \left[-\frac{\pi}{2}, \frac{\pi}{2}
\right]$ with $\vert\vert m \vert \vert \leq 1/2$ and $\supp (m)
\subset [-\pi \vert\vert m \vert \vert,\pi \vert\vert m
\vert\vert]$. In case $||m|| = \half$, assume that there exists
$\delta_0 > 0$ such that for $0 < \delta \leq \delta_0$, \be
m\left( \left\{ q \in [-\frac{\pi}{2},\frac{\pi}{2}]:\,
\frac{\pi}{2} - |q| \leq \delta \right\} \right) \leq
\frac{1}{\pi} \delta. \ee (Notice that the uniform distribution
satisfies this condition.) Let $\Delta = -\cos \mu$ with $\mu
\in (\pi/2,\pi).$ Then there exists a unique continuous function
$k:[-\pi/2,\pi/2] \to [-\pi +\mu, \pi - \mu]$ such that \be k(q)
= q - \int_{-\pi/2}^{\pi/2} \theta (k(q),k(q^\prime)) \,
m(dq^\prime). \label{BAE} \ee
\end{theorem}

{\it Proof.} Define the new function $g (q)$ by \be e^{ik(q)} =
\frac{e^{i\mu} - e^{g (q)}}{e^{i\mu + g (q)} - 1}.
\label{rhodef} \ee

Then $k(q) = K(g(q))$ where $K:\mathbb{R} \to (-\pi/2,\pi/2)$ is
an increasing function given by \be K(\alpha) = \int_0^\alpha
\frac{\sin (\mu)}{\cosh (\beta) - \cos \mu} d\beta = 2\tan^{-1}
\left(\frac{\tanh (\alpha/2)}{\tan (\mu/2)}\right). \label{Kdef} \ee It
follows that $g (q)$ must satisfy: \be K(g (q)) = q -
\int_{-\pi/2}^{\pi/2} \omega (g (q) - g (q^\prime)) \,
m(dq^\prime) \label{BAEtrf} \ee where \be \omega (\alpha) = -2
\tan^{-1} \left( \frac{\tanh (\alpha/2)}{\tan(\mu)} \right).
\label{omegadef} \ee Notice that \be \omega^\prime (\alpha) = -
\frac{\sin (2\mu)}{\cosh \alpha - \cos (2\mu)} > 0.
\label{omegaprime} \ee As in \cite{DLP}, we now define a
functional $B[g]$ on the space $L^2(\RR,m)$ by
\begin{eqnarray} B[g] &=& \int S(g (q)) \, m(dq) - \int
q\, g(q) m(dq) \non \\ && \quad + \frac{1}{2} \int \int \Omega
(g (q) - g (q^\prime))\, m(dq) m(dq^\prime), \label{Bdef}
\end{eqnarray} where $ S(\alpha) = \int_0^\alpha K(\beta) d\beta $
and $\Omega (\alpha ) = \int_0^\alpha \omega (\beta) d\beta$.

The functional $B$ is well-defined because $0 \leq S(\alpha)
\leq \half K'(0) \alpha^2$ and \\ $0 \leq \Omega(\alpha) \leq
\half \omega'(0) \alpha^2$, where $$ K'(0) = \frac{\sin
(\mu)}{1- \cos (\mu)} \mbox{ and } \omega'(0) = -\frac{\sin
(2\mu)}{1-\cos (2\mu)}. $$ It is also easily seen to be
continuous. The Gateaux derivative in the direction of a
function $f$ is given by
\be DB[g]f = \int \bigg\{ K(g (q)) - q
+ \int \omega(g (q) - g(q^\prime)) m(dq^\prime) \bigg\} f(q)
m(dq). \ee It follows that the solution to (3.4) is a stationary
point of $B$. Moreover, $B$ is convex as
\begin{eqnarray} \frac{d^2}{dt^2}
B[g + t f] &=& \int K^\prime (\alpha (q)) f(q)^2 m(dq) \non \\
&& + \half \int\int \omega^\prime (g(q) - g (q^\prime)) (f(q) -
f(q^\prime))^2 m(dq)\, m(dq^\prime) > 0 \non
\\ &&
\end{eqnarray} by (\ref{omegaprime}) and the fact that $K'(\alpha) > 0$.
This proves the uniqueness of the solution.
To prove the existence, we need to find a compact set
which contains the minimiser.

Consider first the case that $||m|| < \half$. Now, as $\alpha \to
\pm \infty$, $ K(\alpha) \to \pm(\pi - \mu)$ and $ \omega (\alpha)
\to \pm (2\mu - \pi)$. Let $M$ be so large that $\pi - \mu -
|K(\alpha)| < \eps$ and $(2\mu - \pi) - |\omega(\alpha)| <  \eps$
for $|\alpha| > M$, where $\eps > 0$ is to be determined later.
Consider the set \be \Gamma_M = \{ q \in [-\pi ||m||, \pi
||m||]:\, g(q) > M\}. \ee For $M$ large enough, we can assume that
$m(\Gamma_M) < \eps.$ We now replace $g$ on the set $\Gamma_{2M}$
by $\pm 2M$, i.e. we set \be {\tilde g}(q) = \mbox{sgn}\,(g(q))
\min\{|g(q)|, 2M\}. \ee By convexity of the functions $\Omega$
and $S$ we then have
\begin{eqnarray} \lefteqn{B[g] - B[{\tilde g}] =} \non \\ &=& \int
(S(g(q)) - S({\tilde g}(q)))\, m(dq) - \int q\,(g(q) - {\tilde
g}(q))\, m(dq) \non \\ && \quad + \half \int \int \left(
\Omega(g(q) - g(q')) - \Omega({\tilde g}(q) - {\tilde g}(q'))
\right) \, m(dq) m(dq') \non \\ &\geq&
\int_{\Gamma_{2M}} (|g(q)| - 2M) (K(2M)-|q|) \, m(dq) \non \\
&& \quad + \int_{\Gamma_{2M}} m(dq) \int_{\Gamma_{2M}^c} m(dq')
\left( \Omega(g(q) - g(q')) - \Omega({\tilde g}(q) - g(q'))
\right) \non \\ && \end{eqnarray} where we used the convexity of
the function $S$ and the fact that if $q,q' \in \Gamma_{2M}$ then
the second term in the double integral is zero whereas the first
term is positive since $\Omega \geq 0$. Next using the convexity
of $\Omega$ and the above bounds on the derivatives we get
\begin{eqnarray}
\lefteqn{B[g] - B[{\tilde g}] =} \non \\ &\geq &
\int_{\Gamma_{2M}} (|g(q)| - 2M) (K(2M)-|q|) \, m(dq) \non \\
&& \quad + \int_{\Gamma_{2M}} m(dq) \int_{\Gamma_{2M}^c} m(dq')
\omega(2M - |g(q')|) (|g(q)|-2M) \non \\ &\geq &
\int_{\Gamma_{2M}} (|g(q)| - 2M) (\pi-\mu-|q|- \eps) m(dq) \non
\\ && \quad + \int_{\Gamma_{2M}} m(dq) \int_{\Gamma_{M}^c} m(dq')
\left( (2\mu - \pi) - \eps \right) (|g(q)|-2M) \non \\ &\geq&
\int_{\Gamma_{2M}} m(dq) (|g(q)| - 2M) \bigl( \pi-\mu-|q|- \eps
+ \left( (2\mu - \pi) -  \eps \right) m(\Gamma_M^c) \bigr) \non \\
&\geq& \int_{\Gamma_{2M}} m(dq) (|g(q)| - 2M)
\bigl( \pi-\mu-|q|- \eps 
+ \left( (2\mu - \pi) - \eps \right) (||m||-\eps) \bigr) \non \\
&\geq& \int_{\Gamma_{2M}} m(dq) (|g(q)| - 2M) \left(
(\pi-\mu)(1-2||m||) -\eps(1+2\mu-\pi+||m||)\right) \non
\\ &>& 0
\end{eqnarray}
provided $$ \eps < \frac{(\pi - \mu)(1-2||m||)}{1+2\mu-\pi+||m||}.
$$
We conclude that the minimiser must satisfy $ ||g||_\infty \leq
2M$ and is a fortiori contained in the ball $\{g \in L^2(m):\,
||g||_2 \leq M\}$. This ball is bounded and therefore weakly
compact. But the functional $B[g]$ is norm continuous and convex
and therefore lower semicontinuous for the weak topology, see e.g.
\cite{Barbu}, Prop. 1.5 of Chap. 2. It follows that it attains its
minimum on a compact set.

\medskip

Next consider the case $||m|| = \half$. In that case we cannot
prove that the minimiser is bounded, so we need a more
sophisticated bound. We use the function $$ f(q) = -2\ln
\left(\frac{\pi}{2}-|q| \right). $$ Given $M>0$ and $\delta > 0$,
we define the sets \be \Gamma_0^M = \{ q \in [-\shalf \pi, \shalf
\pi]:\, |g(q)| > M , |q| < \frac{\pi}{2} - \delta\} \ee and \be
\Gamma_k = \left\{ q \in [-\shalf \pi, \shalf \pi]:\, |g(q)| >
f(q), \frac{\pi}{2} - \gamma^{-k+1} \delta \leq |q| <
\frac{\pi}{2} - \gamma^{-k} \delta \right\}, \ee where $\gamma >
1$ is a parameter to be determined later.

We now write $$\Gamma^M = \Gamma^M_0 \cup \bigcup_{k\geq 1}
\Gamma_k $$ and consider the decomposition \begin{eqnarray}
\lefteqn{\left\{ (q,q') \in [-\shalf \pi,\shalf \pi]^2:\, q \in
\Gamma^M \mbox{ or } q' \in \Gamma^M \right\} =} \non \\ &=&
\bigcup_{k \geq 0} \left( \Gamma_k \times \left( \bigcup_{l \geq
k} \Gamma_l \right)^c \cup \left( \bigcup_{l \geq
k} \Gamma_l \right)^c \times \Gamma_k 
\cup 
(\Gamma_k \times \Gamma_k ) \right). \label{decomp}
\end{eqnarray} Note that this is a disjoint union. Replacing now
$g(q)$ by
$$ {\tilde g}(q) = \mbox{sgn}\,(g(q))\, \min\left\{ |g(q)|
, \left(f(q) \chi_{\Gamma^M\setminus \Gamma^M_0} + 2M
\chi_{\Gamma_0^M} \right) \right\} $$ we have first of all
\begin{eqnarray} \lefteqn{\int (S(g(q)) - S({\tilde
g}(q)))\, m(dq) - \int q\,(g(q) - {\tilde g}(q))\, m(dq)}
\non \\ &\geq& \int_{\Gamma^{2M}_0} m(dq) (|g(q)| - 2M) (K(2M)- |q|) \non \\
&& + \sum_{k=1}^\infty \int_{\Gamma_k} m(dq) \left( |g(q)|
- f(q) \right) \left( K(f(q)) - |q| \right) \non \\
&\geq& \int_{\Gamma^{2M}_0} m(dq) (|g(q)| - 2M) (\pi - \mu - |q|-
\eta) \non \\ && + \sum_{k=1}^\infty \int_{\Gamma_k} m(dq) \left(
|g(q)| - f(q) \right) \non \\ && \qquad \times \left(\pi - \mu -
\frac{\pi}{\tan (\mu/2)} \left( \frac{\pi}{2} - |q| \right)^{2} -
|q| \right), \label{Sbnd}
\end{eqnarray} where we used the bound \be K(f(q)) > \pi - \mu -
\frac{4}{\tan (\mu/2)} \left( \frac{\pi}{2} - |q| \right)^{2} \ee
which follows from the inequalities $$ \tan^{-1}(x-\delta)\ge
\tan^{-1}(x) - \delta $$ and $$ \tanh (x) > 1 - 2 e^{-|x|}. $$

For the term $$ \half \int \int \left( \Omega(g(q) - g(q')) -
\Omega({\tilde g}(q) - {\tilde g}(q')) \right) \, m(dq) m(dq')
$$ we consider the contributions from the decomposition
(\ref{decomp}) separately: \begin{eqnarray} && \int_{\Gamma^M_0}
\int_{(\Gamma^M)^c} \left( \Omega(g(q) - g(q')) - \Omega({\tilde
g}(q) - {\tilde g}(q')) \right) \, m(dq) m(dq') \non \\ && \qquad
\geq \int_{\Gamma^{2M}_0} m(dq) (2\mu - \pi - \eta)
m((\Gamma^M)^c)
\end{eqnarray} as before. Combining this with the first term of
(\ref{Sbnd}) gives a positive contribution provided $M$ is so
large that $m(\Gamma^M) < \eps$ and $K(2M) > \pi - \mu - \eta$ and
$\omega(M) > 2\mu - \pi - \eta$ where $\frac{3}{2} \eta + \pi \eps
< \delta$.

Next consider a term of the form
$$ \int_{\Gamma_k}
\int_{(\cup_{l\geq k} \Gamma_k)^c} \left( \Omega(g(q) - g(q')) -
\Omega({\tilde g}(q) - {\tilde g}(q')) \right) \, m(dq) m(dq').
$$ Assuming $\delta < \delta_0$, this is bounded by
\begin{eqnarray} \lefteqn{\int_{\Gamma_k} \int_{(\cup_{l\geq k}
\Gamma_k)^c} \left( \Omega(g(q) - g(q')) - \Omega({\tilde g}(q) -
{\tilde g}(q')) \right) \, m(dq) m(dq')} \non
\\ &\geq & \int_{\Gamma_k} m(dq) \left( |g(q)| -
f(q) \right) \non \\ && \times \left[ 2\mu - \pi -
\frac{4}{\tan(\pi - \mu)} \left( \frac{\pi}{2} - |q| \right)^{2}
\right] m \left[\left( \bigcup_{l \geq k} \Gamma_l \right)^c
\right].
\end{eqnarray} Since $$ \bigcup_{l \geq k} \Gamma_l \subset
\left\{ q \in \left[ -\frac{\pi}{2}, \frac{\pi}{2} \right]:\,
\frac{\pi}{2} - \gamma^{-k+1} \delta \leq |q| \right\} $$ we have
by the assumption about $m$, \be m \left( \bigcup_{l \geq k}
\Gamma_l \right) \leq \frac{1}{\pi} \gamma^{-k+1} \delta. \ee
Therefore \begin{eqnarray} \lefteqn{\int_{\Gamma_k}
\int_{(\cup_{l\geq k} \Gamma_k)^c} \left( \Omega(g(q) - g(q')) -
\Omega({\tilde g}(q) - {\tilde g}(q')) \right) \, m(dq) m(dq')}
\non
\\ &\geq & \int_{\Gamma_k} m(dq) \left( |g(q)| -
f(q) \right) \non \\ && \times \left[ 2\mu - \pi -
\frac{4}{\tan(\pi - \mu)} \left( \frac{\pi}{2} - |q| \right)^{2}
\right] \left( \half - \frac{1}{\pi} \gamma^{-k+1} \delta \right).
\end{eqnarray} Combining this with the corresponding term of
(\ref{Sbnd}) we have \begin{eqnarray} && \int (S(g(q)) - S({\tilde
g}(q)))\, m(dq) - \int q\,(g(q) - {\tilde g}(q))\, m(dq) \non
\\ && + \int_{\Gamma_k} \int_{(\cup_{l\geq k} \Gamma_k)^c} \left(
\Omega(g(q) - g(q')) - \Omega({\tilde g}(q) - {\tilde g}(q'))
\right) \, m(dq) m(dq') \non \\ && \quad \geq \int_{\Gamma_k}
m(dq) \left( |g(q)| - f(q) \right) \non \\ && \qquad \times \left[
\pi - \mu - \frac{\pi}{\tan (\mu/2)} \left( \frac{\pi}{2} - |q|
\right)^{2} - |q| \right. \non
\\ && \qquad + \left. \left( 2\mu - \pi - \frac{4}{\tan(\pi -
\mu)} \left( \frac{\pi}{2} - |q| \right)^{2} \right) \left( \half
- \frac{1}{\pi} \gamma^{-k+1} \delta \right) \right].
\end{eqnarray} Since $\frac{\pi}{2} - \gamma^{-k+1} \delta \leq
|q| < \frac{\pi}{2} - \gamma^{-k} \delta$ for $q \in \Gamma_k$, we
have
\begin{eqnarray} && \pi - \mu - c_1 \left( \frac{\pi}{2} - |q|
\right)^2 - |q| \non \\ && + \left( 2\mu - \pi - c_2 \left(
\frac{\pi}{2} - |q| \right)^2 \right) \left( \half - \frac{1}{\pi}
\gamma^{-k+1} \delta \right) \non \\ && \quad \geq \frac{\pi}{2} -
\mu - c_1 \left( \gamma^{-k+1} \delta \right)^2 + \gamma^{-k}
\delta \non \\ && \qquad + \mu - \frac{\pi}{2} - \half c_2 \left(
\gamma^{-k+1} \delta \right)^2 - \frac{2\mu - \pi}{\pi}
\gamma^{-k+1} \delta \non \\ \quad  &=& \left( 1 - \frac{ 2\mu -
\pi}{\pi} \gamma \right) \gamma^{-k} \delta - c \gamma^{-2k+2}
\delta^2, \end{eqnarray} where $$ c_1 = \frac{4}{\tan(\mu/2)},
\quad c_2 = \frac{4}{\tan(\pi - \mu)}, \mbox{ and } c = c_1 +
\half c_2. $$ Hence
\begin{eqnarray} && \int (S(g(q)) -
S({\tilde g}(q)))\, m(dq) - \int q\,(g(q) - {\tilde g}(q))\, m(dq)
\non \\ && + \int_{\Gamma_k} \int_{(\cup_{l\geq k} \Gamma_k)^c}
\left( \Omega(g(q) - g(q')) - \Omega({\tilde g}(q) - {\tilde
g}(q')) \right) \, m(dq) m(dq') \non \\ && \quad \geq
\int_{\Gamma_k} m(dq) \left( |g(q)| - f(q) \right) \non \\ &&
\qquad \times \left[ \left( 1 - \frac{ 2\mu - \pi}{\pi} \gamma
\right) \gamma^{-k} \delta - c \gamma^{-2k+2} \delta^2 \right].
\end{eqnarray} Finally consider the terms $$ \half \int_{\Gamma_k}
m(dq) \int_{\Gamma_k} m(dq') \left( \Omega (g(q)-g(q')) -
\Omega(f(q)-f(q')) \right). $$ Since $0 \leq \Omega(\alpha) \leq
(2\mu -\pi) |\alpha|$, these can be bounded by
\begin{eqnarray} && \half \int_{\Gamma_k} m(dq) \int_{\Gamma_k}
m(dq') \left( \Omega (g(q)-g(q')) - \Omega(f(q)-f(q')) \right)
\non \\ && \quad \geq -(\mu - \frac{\pi}{2}) \int_{\Gamma_k} m(dq)
\int_{\Gamma_k} m(dq') \, |f(q) - f(q')| \non \\ && \quad \geq \pi
\ln (\gamma^{-k} \delta) m(\Gamma_k)^2 \geq \pi
(\gamma^{-k}\delta)^2 \ln (\gamma^{-k} \delta).
\end{eqnarray} In all, we get \begin{eqnarray} && \int (S(g(q))
- S({\tilde g}(q)))\, m(dq) - \int q\,(g(q) - {\tilde g}(q))\,
m(dq) \non \\ && + \int_{\Gamma_k} \int_{(\cup_{l\geq k}
\Gamma_k)^c} \left( \Omega(g(q) - g(q')) - \Omega({\tilde g}(q) -
{\tilde g}(q')) \right) \, m(dq) m(dq') \non \\ && + \half
\int_{\Gamma_k} m(dq) \int_{\Gamma_k} m(dq') \left( \Omega
(g(q)-g(q')) - \Omega(f(q)-f(q')) \right) \non \\ && \quad
\geq \int_{\Gamma_k} m(dq) \left( |g(q)| - f(q) \right) \non \\
&& \qquad \times \left[ \left( 1 - \frac{ 2\mu - \pi}{\pi} \gamma
\right) \gamma^{-k} \delta - c \gamma^{-2k+2} \delta^2  + \pi
(\gamma^{-k}\delta)^2 \ln (\gamma^{-k} \delta) \right].
\end{eqnarray} Choosing $\gamma < \frac{2\mu - \pi}{\pi}$ (which
is possible as $\mu < \pi$) and $\delta $ small enough, this is
positive.

It now follows that in this case the minimiser of $B[g]$ must
satisfy
$$ |g(q)| \leq f(q) \chi_{[-\half \pi, -\half \pi + \delta] \cup
[\half \pi - \delta, \half \pi]} +
2M \chi_{[-\half \pi + \delta, \half \pi - \delta]} $$ and therefore
\begin{eqnarray} ||g||_2^2 &\leq& 2M^2 + 2\int_{\half \pi-\delta}^{\half \pi}
f(q)^2 m(dq) \non \\ &\leq& 2M^2 + \frac{8}{\pi} \int_0^\delta
(\ln x)^2 dx < +\infty.
\end{eqnarray} Again, it follows that $B[g]$ attains its minimum on
this compact set.

We finally prove that the unique solution $g \in L^2(\RR,m)$ of
(\ref{BAEtrf}) in fact has a continuous version as a function $g:
[-\pi ||m||,\pi ||m||]  \to [-\infty,\infty]$. We have shown that
there exists ${\tilde g} \in {\cal L}^2(\RR,m)$ satisfying
(\ref{BAEtrf}) for $m$-a.e. $q$. We define the image measure
${\tilde m} = {\tilde g}(m)$ and put \begin{equation} h(x) = K(x)
+ \int \omega(x-x') {\tilde m}(dx).
\end{equation} Clearly, $h$ is a ${\cal C}^\infty$-function on $\RR$ and
$$ h'(x) = K'(x) + \int \omega'(x-x') {\tilde m}(dx) > 0. $$
Therefore, the inverse function $g = h^{-1}$ is well-defined and
${\cal C}^\infty$ on the range of $h$. Since $$ h(x) \to \pm (\pi
- \mu + (2 \mu - \pi) ||{\tilde m}||) $$ as $x \to \pm \infty$,
the function $g$ is defined on the interval
$$ I_m = \bigl( -(\pi-\mu +(2\mu-\pi)||m||),\pi-\mu + (2\mu-\pi)||m|| \bigr). $$
Notice that if $||m|| < \half$, this interval contains $[-\pi
||m||,\pi |m||]$, whereas if $||m|| = \half$, $I_m = (-\shalf \pi,
\shalf \pi)$. In the latter case, $g$ extends continuously as a
function $g:\,[-\shalf \pi, \shalf \pi] \to [-\infty,+\infty]$.
Inserting $x = {\tilde g}(q)$ we have for $q \in
\mbox{supp}\,(m)$,
$$ h({\tilde g}(q)) = K({\tilde g}(q)) + \int \omega({\tilde g}(q) -
{\tilde g}(q'))\,m(dq') = q $$ for $m$-a.e. $q$. Hence $g(q) =
{\tilde g}(q)$ for $m$-a.e. $q$. Now inserting $x=g(q)$ we get
\begin{eqnarray*} q = h(g(q)) &=& K(g(q)) + \int \omega(g(q) -
{\tilde g}(q')) m(dq') \\ &=& K(g(q)) + \int \omega(g(q) - g(q'))
m(dq'),
\end{eqnarray*}
so that $g$ satisfies (\ref{BAEtrf}) for all $q$ in its domain. It
remains to show that the solution $g$ is unique. It follows from
the mean-value theorem that any continuous solution is
differentiable and its derivative is given by
\begin{eqnarray} \label{rhoder} g'(q) &=& \frac{1}{K'(g(q)) +
\int \omega'(g(q)-g(q')) \,m(dq')} \non \\ &=& \frac{1}{K'(g(q)) +
\int \omega'(g(q)-{\tilde g}(q')) \,m(dq')}. \end{eqnarray} Note
that the function ${\tilde g}$ is uniquely defined modulo an
$m$-null-set, so that the right-hand side only depends on the
value of $g$ at $q$. Since $g(q)$ is uniquely defined on
$\mbox{supp}\,(m)$ by continuity, its extension to $[-\pi
||m||,\pi ||m||]$ is also unique.

\qed

\begin{theorem} The mapping $m \mapsto k_m$ defined by (\ref{BAE})
in Theorem~\ref{thm1} is continuous, that is, if $m_n \to m$
weakly then $k_{m_n} \to k_m$ in norm. \label{thm2} \end{theorem}

\textbf{Proof.}  Let $m^{(1)}_n$ be a subsequence. Notice that
$\vert\vert k_{m_n} \vert\vert \leq \pi - \mu$ and $k_{m_n}$ is
also equicontinuous because \begin{equation} \frac{\partial }{
\partial k} \theta (k,k^\prime) = \Delta \frac{\cos (k^\prime) + \cos
(\mu)}{\Delta^2 \sin^2 (k-k^\prime)/2 + [\cos (k+k^\prime)/2 -
\Delta \cos (k-k^\prime)/2 ]^2} \geq 0
\end{equation} for $-\pi + \mu \leq k^\prime \leq \pi - \mu$.
Hence \begin{equation} k_{m_n}^\prime (q) = \left\{ 1 + \int
\frac{\partial \theta}{\partial k} (k_{m_n} (q) - k_{m_n}
(q^\prime)) m_n (dq^\prime) \right\}^{-1} \in (0,1)
\end{equation} Therefore, $\vert k_{m_n} (q) - k_{m_n}
(q^\prime) \vert \leq \vert q - q^\prime \vert $ uniformly in
$n$. It follows that there exists a subsequence $m_n^{(2)}$ of
$m_n^{(1)}$ such that $k_{m_n^{(2)}}$ converges to a continuous
function $k$ uniformly on $[-\pi/2, \pi/2]$. We must show that
$k=k_m$. But $\theta$ is uniformly continuous on $[-\pi+\mu, \pi
- \mu]^2$ so $\theta (k_{m_n^{(2)}} (q) - k_{m_n^{(2)}} (\cdot))
\to \theta (k(q) - k(\cdot))$ in norm, and hence
$$ \int \theta (k_{m_n^{(2)}} (q) - k_{m_n^{(2)}} (q^\prime)) m_n^{(2)} (dq^\prime) \to
\int \theta (k(q) - k(q^\prime)) m(dq^\prime). $$ It follows that
$k(q) = k_m(q)$.  \qed

\begin{corollary} If $m_n \to m$ weakly, and $\tilde m_n$ is the image
measure of $m_n$ under the mapping $k_{m_n}$ then $\tilde m_n \to
\tilde m = k_m (m)$. \end{corollary}

\textbf{Proof.} Let $F \in {\cal C} ([-\pi + \mu, \pi - \mu])$.
Then $\int F(k) \tilde m_n (dk) = \\ \int F(k_{m_n}(q)) m_n (dq)
$ and \begin{eqnarray} \lefteqn{ \left\vert \int F(k) {\tilde
m}_n (dk) - \int F(k) {\tilde m} (dk) \right\vert \leq } \non \\
&\leq& \int \vert F(k_{m_n}(q)) - F(k_m(q)) \vert m_n (dq) \non \\
&& \qquad + \left\vert \int F(k_m(q)) m_n (dq) - \int F(k_m(q))
m(dq) \right\vert .
\end{eqnarray} The right-hand side tends to zero as $n\to \infty $
because $k_{m_n} \to k_m$ uniformly and $k_m$ is continuous. \qed

\begin{theorem} \label{thm3}  Let
\be m_N = \frac{1}{N} \sum_{j=1}^{n_N} \delta_{q_j}, \ee where $
q_j = \frac{2\pi}{N} (j-\half (n_N +1))$ and $n_N \leq N/2$.
Assume that $n_N/N \to \rho $ as $N \to \infty$. Then $m_N \to
\frac{1}{2\pi} dq$ on $[-\pi\rho , \pi\rho]$ and $ \tilde m_n
\to \tilde m$, where $\tilde m$ is absolutely continuous with
respect to the Lebesgue measure and symmetric, and there exists
$Q \in [0,\pi - \mu]$ such that $\supp({\tilde m}) = [-Q,Q]$.
\end{theorem}

\textbf{Proof.} Let $F \in {\cal C} ([-\pi/2, \pi/2])$. Then
\begin{equation} \int F(q) \, m_N (dq) = \frac{1}{N} \sum_{j=1}^{n_N} F
\left( \frac{2\pi}{N} (i - \half (n_N + 1)) \right)  \to
\int_{-\pi\rho}^{\pi\rho} F(q) \frac{dq}{2\pi}. \end{equation}
It follows that $\tilde m_N \to \tilde m$, and we must show that
$\tilde m$ is absolutely continuous and even. The latter follows
from the fact that $k_m$ is even, which is a consequence of the
uniqueness. To prove the absolute continuity, let $\eps > 0$. We
must show that there exists $\delta > 0$ such that $\tilde m
(k_0 - \delta , k_0 + \delta) < \eps$ for all $k_0$. Now,
$\tilde m (k_0+\delta, k_0+\delta) = \int_{k_m^{-1} (k_0-\delta,
k_0+ \delta)} \frac{dq}{2\pi}$ and we have seen that $k_m$ is
continuous and increasing: $[-\pi/2,\pi/2] \to [-\pi+\mu, \pi -
\mu]$. Therefore $k_m^{-1}$ is continuous and $\forall \eps
>0 \, \exists \delta > 0:\, k_m^{-1} (k_0-\delta, k_0+\delta)
\subset (q_0 - \pi\eps, q_0+\pi\eps)$ where $k_m (k_0) = q_0$.
Hence $\int_{k_m^{-1} (k_0-\delta, k_0+\delta)} \frac{dq}{2\pi}
< \eps.$ \qed

Writing the free energy (\ref{BAfe}) in the form
\begin{eqnarray} f(\beta, \rho) &=& \lim_{N \to \infty}
\min \left\{\eps_1 - \frac{1}{\beta} \int \ln L(e^{ik(q)}) \,
m_N(dq), \right. \non \\ && \qquad\qquad \qquad  \left. \eps_2 -
\frac{1}{\beta} \int \ln M(e^{ik(q)})\, m_N(dq) \right\} \non \\
&=& \lim_{N \to \infty} \min \left\{\eps_1 - \frac{1}{\beta} \int
\ln L(e^{ik}) \, {\tilde m}_N(dk), \right. \non \\ && \qquad
\qquad \qquad \left. \eps_2 - \frac{1}{\beta} \int \ln M(e^{ik})\,
{\tilde m}_N(dk) \right\}
\end{eqnarray} we obtain \begin{eqnarray} \label{fmin}
f(\beta, \rho) &=& \min \left\{\eps_1 - \frac{1}{2 \pi \beta}
\int \ln L(e^{ik(q)}) \, dq, \eps_2 - \frac{1}{2 \pi \beta} \int
\ln M(e^{ik(q)})\, dq \right\} \non \\ &=& \min \left\{\eps_1 -
\frac{1}{\beta} \int \ln
L(e^{ik}) \, {\tilde m}(dk), 
\eps_2 - \frac{1}{\beta} \int \ln M(e^{ik})\, {\tilde m}(dk)
\right\}\!. \non \\ &&
\end{eqnarray} By transformation to the variable $y = g(q)$ this
becomes
\begin{eqnarray} f(\eps_1,\eps_2,\eps_3;\beta) &=&
\min \left\{\eps_1 - \frac{1}{2 \pi\beta}
\int_{-y_1}^{y_1} \ln L(e^{iK(y)}) R(y) dy, \right. \non \\
&& \qquad \left. \eps_2 - \frac{1}{2 \pi \beta} \int_{-y_1}^{y_1}
\ln M(e^{iK(y)}) R(y) dy \right\}.
\end{eqnarray} Here $y_1 = g(\pi/2)$ and
$R(y) = g'(q)^{-1}$ is given by (\ref{rhoder}):
\begin{eqnarray} R(y) &=& K'(y) + \int_{-\pi/2}^{\pi/2}
\omega'(y-g(q)) \frac{dq}{2 \pi} \non \\ &=& \frac{\sin
(\mu)}{\cosh (y) - \cos (\mu)} - \frac{1}{2\pi} \int_{-y_1}^{y_1}
\frac{\sin(2\mu)}{\cosh(y-\alpha) - \cos(2 \mu)} R(\alpha)
d\alpha. \non \\ &&
\end{eqnarray} If we assume that $y_1 = +\infty$ then this can be
evaluated by Fourier transformation as in \cite{Baxter}. With
\begin{equation} {\hat R}(x) = \frac{1}{2\pi} \int_{-\infty}^\infty
R(\alpha) e^{i\alpha x} d\alpha \end{equation} we have $ {\hat
R}(x) = e^{-\mu x} - e^{-2\mu x} {\hat R}(x)$ and hence
\begin{equation} {\hat R}(x) = \frac{1}{2 \cosh (\mu x)}.
\end{equation} This is consistent with the fact that the minimum
in (\ref{fmin}) is attained at $\rho = \half$ since $f$ is
convex in $\rho$. Indeed, $$ {\hat R}(0) =  \frac{1}{2\pi}
\int_{-y_1}^{y_1} R(\alpha) d\alpha = ||m|| = \rho. $$ Now,
$R(y) = (g^{-1})'(y)$ determines $g^{-1}$ given that $g^{-1}(0)
= 0$. By uniqueness of $g(q)$, it must be the solution.
Eventually, one finds

\begin{corollary} Assume $\Delta \in [0,1)$. Then, the free energy
is given by
\begin{eqnarray} f(\beta) &=& \eps_1 - \frac{1}{\beta}
\int_{-\infty}^\infty \frac{\sinh[(\mu +
w)x]\,\sinh[(\pi-\mu)x]}{2 x \sinh[\pi x]\, \cosh[\mu x]} dx \non
\\ &=& \eps_2 - \frac{1}{\beta}
\int_{-\infty}^\infty \frac{\sinh[(\mu - w)x]
\,\sinh[(\pi-\mu)x]}{2 x \sinh[\pi x]\, \cosh[\mu x]} dx,
\end{eqnarray} where the parameter $w$ is defined by
\begin{equation} a\,:\,b\,:\,c = \sin \shalf (\mu - w) \,:\,
\sin \shalf (\mu + w) \,:\, \sin(\mu). \end{equation}
\end{corollary}

\section{The case $\Delta \ll -1$.}
\setcounter{equation}{0}

For $\Delta < -1$ we write $\Delta = -\cosh \lambda$, assuming
$\lambda > 0$. Define the new function $\alpha (q)$ by \be
e^{ik(q)} = \frac{e^{\lambda} - e^{-i \alpha(q)}}{e^{\lambda-i
\alpha (q)} - 1}. \label{alphadef} \ee Then $k(q) =
\omega_\lambda(\alpha(q))$ where $\omega_\lambda:[-\pi,\pi] \to
[-\pi,\pi]$ is an increasing function given by \be
\omega_\lambda(x) = \int_0^x \frac{\sinh (\lambda)}{\cosh
(\lambda) - \cos (u)} du = 2 \tan^{-1}\left(\frac{\tan (x/2)}{\tanh
(\lambda/2)}\right). \label{omegadef2} \ee In terms of $\alpha(q)$ the
Bethe Ansatz equations read \be  \omega_\lambda(\alpha (q)) = q
+ \int_{-\pi/2}^{\pi/2} \omega_{2\lambda} (\alpha (q) - \alpha
(q^\prime)) \, m(dq^\prime). \label{BAEtrf2} \ee Here the
function $\omega_{2\lambda}$ is defined as in (\ref{omegadef2})
with the understanding that for $|x| > \pi$ the integral
expression is assumed so that $\omega_{2\lambda}$ is continuous.
The measure $m \in {\cal M}^b_+ \left[
-\frac{\pi}{2},\frac{\pi}{2} \right]$ satisfies $||m|| \leq
\frac{1}{2}$. For the finite lattice, it is given by \be m =
\frac{1}{N} \sum_{j=1}^n \delta_{q_j}, \qquad q_j =
-\frac{n+1-2j}{2 n}\pi. \ee

\begin{theorem} \label{thm5}
Assume $\lambda > \lambda_0$, where $\lambda_0 =\ln (3 + 2\sqrt{5}),$  when $||m||\le 1/2$. Then, for any measure $m \in {\cal M}^b_+
[-\frac{\pi}{2},\frac{\pi}{2}]$ with $||m|| \leq \half$, there
exists a unique function $\alpha \in L^\infty(m)$ such that
(\ref{BAEtrf2}) holds for all $q \in \mbox{\rm supp}\,(m)$.
Moreover, if $m$ is symmetric then $\alpha$ extends uniquely to a
continuous function on $[-\pi/2,\pi/2]$ with values in
$[-\pi,\pi]$ which satisfies (\ref{BAEtrf2}) for all $q \in
[-\pi/2,\pi/2]$.
\end{theorem}

\textbf{Proof.} We expand $\tan^{-1}\left(\tan(x/2)/\tanh(\lambda/2)\right)$  into a Fourier
series. For this, we compute first  $$ \frac{1}{2\pi}
\int_{-\pi}^\pi \frac{\sinh(\lambda)}{\cosh(\lambda)-\cos(x)}
e^{inx} dx = e^{-\lambda |n|}. $$ Hence $$
\frac{\sinh(\lambda)}{\cosh(\lambda) - \cos(x)} = \sum_{n \in \ZZ}
e^{-\lambda |n|} e^{-inx} = 1 + 2 \sum_{n=1}^\infty e^{-\lambda n}
\cos(nx).$$ Integrating, we obtain the following Fourier
expansion: \be \omega_\lambda(x) = x + 2 \sum_{n=1}^\infty
\frac{e^{- \lambda n}}{n} \sin(n x). \ee Inserting this into the
Bethe Ansatz equation (\ref{BAEtrf2}), we have \be
\begin{array}{r}
\displaystyle\alpha(q)=
\displaystyle \frac{q}{1-||m||} -\frac{1}{1-||m||}\int\alpha(q')m(dq')-
\displaystyle\frac{2}{1-||m||}\sum_{n=1}^{\infty}
e^{-n\lambda}\frac{\sin(n\alpha(q))}{n}+\label{FourierBAE}\\[4ex]
+\displaystyle\frac{2}{1-||m||} \sum_{n=1}^\infty e^{-2\lambda n}\int \frac{\sin(n(\alpha(q)-\alpha(q')))}{n}m(dq'). 
\end{array}
\ee
Notice next that it follows from (\ref{FourierBAE}):
$$
\int\alpha(q')m(dq')=\int q'm(dq')-2\sum_{n=1}^{\infty}\frac{1}{n}\int e^{-n\lambda}\sin(n\alpha(q'))m(dq').
$$ 
Let us introduce an operator:
$$
\begin{array}{l}
T[\alpha]=\\[2ex]
-\displaystyle\frac{1}{1-||m||}\int q'm(dq')+\frac{2}{1-||m||}\sum_{n=1}^{\infty}\frac{1}{n}\int e^{-n\lambda}\sin(n\alpha(q'))m(dq')
+\displaystyle \frac{q}{1-||m||}-\\[3ex]
-\displaystyle\frac{2}{1-||m||}\sum_{n=1}^{\infty}
e^{-n\lambda}\frac{\sin(n\alpha(q))}{n}+\displaystyle\frac{2}{1-||m||} \sum_{n=1}^\infty e^{-2\lambda n}\int \frac{\sin(n(\alpha(q)-\alpha(q')))}{n}m(dq').
\end{array}
$$
We consider (\ref{FourierBAE}) as a fixed point problem 
$\alpha=T[\alpha]$ and show that the map $\alpha \mapsto T[\alpha]$ is contraction w.r.t. the $L^\infty$ norm for sufficiently large $\lambda$.
This is straightforward:
\begin{eqnarray*}
||T(\alpha_1)-T(\alpha_2)||\leq
\left(\frac{1+||m||}{1-||m||}\frac{2}{e^{\lambda}-1}+
\frac{4||m||}{1-||m||}\frac{1}{e^{2\lambda}-1}\right)
||\alpha_{1}-\alpha_{2}||.
\end{eqnarray*}
Clearly, the factor in front of
$||\alpha_1-\alpha_2||$ is less than 1 if $\lambda > \ln (3 + 2\sqrt{5})$. 

The same inequalitie holds for functions $\alpha \in {\cal
C}([-\frac{\pi}{2},\frac{\pi}{2}])$, so there also exists a
unique continuous function satisfying (\ref{BAEtrf2}) for all $q
\in [-\frac{\pi}{2},\frac{\pi}{2}]$. Clearly, restricting this
function to ${\rm \supp}(m)$ yields the solution $\alpha \in
L^\infty(m)$. Finally, note that if $m$ is symmetric, uniqueness
implies that the function $\alpha$ must be odd. This in turn
implies that $\alpha(q) \in [-\pi,\pi]$ for if $\alpha_1\,\alpha_2
\in [0,\pi]$ then $$ \omega_{2\lambda}(\alpha_1-\alpha_2) +
\omega_{2\lambda}(\alpha_1+\alpha_2) \leq 2
\omega_{2\lambda}(\alpha_1) $$ as follows easily by
differentiation. \qed

We now also have an analogue of Theorem~\ref{thm3}:
\begin{lemma} The map $m \mapsto \alpha_m$ defined by
Theorem~\ref{thm5} is continuous. \end{lemma}

\textbf{Proof.} This follows from (\ref{FourierBAE}) using
$$
\begin{array}{l}
 \displaystyle|\int\sin(n\alpha_{m_1}(q'))m_{1}(dq') - \int\sin(n \alpha_{m_2}(q'))m_{2}(dq')|,\\[2ex]
 \displaystyle|\int\cos(n\alpha_{m_1}(q'))m_{1}(dq') - \int\cos(n \alpha_{m_2}(q'))m_{2}(dq')| \leq \half n ||m_1-m_2||. 
\end{array} 
$$ 
\qed

\begin{corollary} If $m_n \to m$ weakly, and $\tilde m_n$ is the image
measure of $m_n$ under the mapping $k_{m_n}$ then $\tilde m_n \to
\tilde m = k_m (m)$. \end{corollary}

As in the case $\Delta \in [0,1)$ we can now conclude that the
free energy is given by
\begin{eqnarray} f(\beta, \rho)
&=& \min \left\{\eps_1 - \frac{1}{2 \pi \beta} \int \ln
L(e^{ik(q)}) \, dq, \right. \non \\ && \qquad \left. \eps_2 -
\frac{1}{2 \pi \beta} \int \ln M(e^{ik})\, dq \right\}.
\end{eqnarray} Transforming to the variable $\alpha$ we have
\begin{eqnarray} f(\eps_1,\eps_2,\eps_3;\beta) &=&
\min \left\{\eps_1 - \frac{1}{2 \pi\beta} \int_{-\pi}^\pi
\ln L(e^{i\omega_\lambda(\alpha)}) R(\alpha) d\alpha, \right. \non
\\ && \qquad \left. \eps_2 - \frac{1}{2 \pi \beta}
\int_{-\pi}^\pi \ln M(e^{i\omega_\lambda(\alpha)})
R(\alpha) d\alpha \right\},
\end{eqnarray} where $R(\alpha) = \alpha'(q)^{-1}$. Again, it can
be evaluated by Fourier transformation, but now on $[-\pi,\pi]$:
\begin{equation} {\hat R}_p = \frac{1}{2 \cosh (\lambda p)}.
\end{equation}

The resulting free energy is
\begin{corollary} Assume $\Delta = -\cosh(\lambda)$ with $\lambda > \lambda_0$.
Then, the free energy of the 6-vertex model with periodic boundary
conditions is given by
\begin{eqnarray} f(\beta) &=& \eps_1 - \frac{1}{\beta}
\left\{ \frac{\lambda + v}{2} + \sum_{p=1}^\infty
\frac{\sinh[(\lambda + v)p]\,e^{-p\lambda}}{p \cosh[p \lambda]}
\right\} \non
\\ &=& \eps_2 - \frac{1}{\beta}
\left\{ \frac{\lambda - v}{2} + \sum_{p=1}^\infty
\frac{\sinh[(\lambda - v)p]\,e^{-p\lambda}}{p \cosh[p \lambda]}
\right\},
\end{eqnarray} where the parameter $v$ is given by
\begin{equation} a\,:\,b\,:\, c = \sinh \shalf (\lambda-v) \,:\,
\sinh \shalf (\lambda+v) \,:\, \sinh(\lambda).
\end{equation} \end{corollary}

\section{Concluding remarks.}

For values of $\Delta<-1$ which are not large negative, numerical
iteration of the equations (\ref{BAEtrf2}) with $m=m_N$ seems to
indicate that there is in fact a unique solution. We have so far
not been able to prove this, although it is possible to show that
the solution is unique and increasing for small $|q|$.


\begin{thebibliography}{99}

\bibitem{Lieb1} E. H. Lieb: Residual Entropy of Square Ice. \emph{Phys.
Rev.} {\bf 162}, 162--172 (1967).

\bibitem{Lieb2} E. H. Lieb: Exact solution of the F-model of an
anti-ferroelectric. \emph{Phys. Rev. Letters} {\bf 18}, 692--694
(1967).

\bibitem{Lieb3} E. H. Lieb: Exact solution of the two-dimensional
Slater KDP model of a ferroelectric. \emph{Phys. Rev. Letters}
{\bf 19}, 108--110 (1967).

\bibitem{Su} B. Sutherland: Exact solution of the two-dimensional
model for hydrogen-bonded crystals. \emph{Phys. Rev. Letters} {\bf
19}, 103--104 (1967).

\bibitem{Baxter} R. J. Baxter: {\sl Exactly Solved Models in
Statistical Mechanics.} Academic Press, 1982.

\bibitem{LiebWu} E. H. Lieb \& F. Y. Wu (with R. J. Baxter):
Two-dimensional Ferroelectric Models. {\bf In:} {\sl Phase
Transitions and Critical Phenomena I.} Eds. C. Domb \& M. S.
Green. Academic Press, 1972. Pp. 331--490.

\bibitem{Brascamp} H. J. Brascamp, H. Kunz \& F. Y. Wu: Some
rigorous results for the vertex model in statistical mechanics.
\emph{J. Math. Phys.} {\bf 14}, 1927--1932 (1973).

\bibitem{DLP} T. C. Dorlas, J. T. Lewis \& J. V. Pul\'e: The Yang-Yang
Thermodynamic Formalism and Large Deviations. \emph{Commun. Math.
Phys.} {\bf 124}, 365--402 (1989).

\bibitem{YY1} C. N. Yang \& C. P. Yang: One-Dimensional Chain of
Anisotropic Spin-Spin Interactions. I. Proof of Bethe's Hypothesis
for GroundState in a FInite System. \emph{Phys. Rev.} {\bf 150},
321--327  (1966).

\bibitem{Bleher} P. Bleher \& K. Liechty: Exact Solution of the
Six-Vertex Model with Domain-Wall Boundary Conditions. Critical
Line between Ferroelectric and Disordered Phases. \emph{J. Stat.
Phys.} {\bf 134}, 463--485 (2009).

\bibitem{Bethe} H. Bethe: Zur Theorie der Metalle I. Eigenwerte
und Eigenfunktionene der linearen Atomkette. Zeits. f. Phys.
\textbf{71}, 205--226 (1931).

\bibitem{Barbu} V. Barbu \& Th. Precupanu, {\sl Convexity and
Optimization in Banach Spaces.} Romania: Editura Academiei, 1978.

\end{thebibliography}
\end{document}